\documentstyle[psfig]{elsart}

\begin{document}
\begin{frontmatter}
\title{Squeezed condensate of gluons and $\eta-\eta'$ mass difference}

\author[rostock1]{D. Blaschke}
\author[rostock2]{H.-P. Pavel\thanksref{DFG}}
\author[dubna]{V.N. Pervushin\thanksref{RFFI}}
\author[rostock1,dubna]{ G. R\"opke} and 
\author[dubna]{M.K. Volkov\thanksref{RFFI}}
\address[rostock1]{MPG Arbeitsgruppe  ''Theoretische Vielteilchenphysik''\\
         Universit\"at Rostock, D-18051 Rostock, Germany}
\address[rostock2]{Fachbereich Physik, Universit\"at Rostock, D-18051 Rostock,
Germany}
\address[dubna]{Bogoliubov  Laboratory of Theoretical Physics,\\
        Joint Institute for Nuclear Research, 141980, Dubna, Russia}
\thanks[DFG]{Supported by DFG grant No RO 905/11-1}
\thanks[RFFI]{Supported by RFFI grant No 96-01-01223}
\noindent MPG--VT--UR 89/96\\
\noindent November 1996

\begin{abstract}
We consider a mechanism to create the $\eta - \eta'$ mass difference 
by the gluon anomaly in a squeezed vacuum.
We find that the mass shift of the $\eta_0$ governing this mass difference
is determined by the magnetic part of the gluon condensate. 
For the squeezed vacuum this magnetic part  
coincides with the total gluon condensate, so that
we get a relation between the gluon condensate and the mass shift
of the $\eta_0$ as a function of the strong coupling constant $\alpha_s$.
The values of the gluon condensate obtained through this relation are
compared with the value by Shifman, Vainshtein and Zakharov
and the recent update values by Narison.
\vspace{5mm}
\\
\noindent PACS number(s):  12.38.Aw, 12.40.Yx, 14.40.Aq, 14.70.Dj
\begin{keyword}
Squeezed vacuum, gluon condensate, U$_A$(1) symmetry breaking
\end{keyword}
\end{abstract}
\end{frontmatter}
\newpage

{\bf 1.} In this note we point out that under the assumption of a squeezed
gluon vacuum a relation
between the $\eta-\eta'$ meson mass difference and the gluon condensate
can be obtained nonperturbatively.
Comparison of the condensate values found through this relation
 with the value of the gluon condensate 
by Shifman, Vainshtein and Zakharov (SVZ) \cite{zakharov}
and with the update average value by Narison \cite{narison}
allows us to check the applicability of the model of a squeezed
gluon vacuum.

Recall that the 
U(1) problem \cite{weinberg} is the question why the $\eta'$ mass is much 
larger than that of the other eight pseudoscalar mesons, especially the $\eta$.
The $\eta-\eta'$ mass difference $m_{\eta '}^2-m_{\eta }^2=0.616$ GeV$^2$
is governed by the mass splitting between the singlet 
and the octet pseudoscalars $\eta_0$ and $\eta_8$, which are related to 
the physical states $\eta, \eta'$ via the mixing

\begin{eqnarray}
\eta &=& \eta_0 \sin\phi - \eta_8\cos\phi\nonumber\\ 
\eta' &=& \eta_0 \cos\phi + \eta_8\sin\phi~. 
\end{eqnarray}

A mixing angle $\phi=-(17\pm 2)^o$ has been obtained in
recent analyses of $\eta $ and $\eta'$ decays \cite{ball,pham}. 
The mass of the $\eta_8$ meson as a member of the pseudoscalar flavour octet
is well explained by explicit chiral symmetry breaking 
in accordance with the Goldstone theorem and the Gell-Mann--Oakes--Renner 
relation.
However, explicit chiral symmetry breaking is not sufficient to explain the 
large mass of $\eta_0$ \cite{weinberg}.

In the literature \cite{thooft}-\cite{volkov} the large mass of $\eta_0$
is explained by the gluon anomaly 
$G^{\mu\nu a} \tilde{G}_{\mu\nu}^a \equiv \partial_{\mu}K^{\mu}$.
There are several ways to implement this gluon anomaly.
In Ref. \cite{thooft} t'Hooft relates this term to the instanton density
in Euclidean space and introduces
an effective quark interaction simulating the anomalous term  
which breaks $U_A(1)$ but conserves
the chiral $SU(3)_L\otimes SU(3)_R$ symmetry. This determinant interaction
has been widely used within effective quark models such as the NJL model
\cite{weise,dmitra}.

Not using the concept of instantons, other authors 
\cite{crew}-\cite{volkov}  start from an effective hadron
Lagrangian which explicitely includes an anomalous meson-gluon interaction 
term which can be viewed in analogy to the anomalous 
$\pi^0 \rightarrow 2 \gamma$ decay  
\begin{equation}
\label{leta}
{\cal L}_{\rm singlet}^{\rm meson}=
\frac{1}{2}\partial_\mu\eta_0\partial^\mu\eta_0
- \eta_0\frac{c}{4}G^{\mu\nu a} \tilde{G}_{\mu\nu}^a~,  
\end{equation} 
where 
\begin{equation}
c= \sqrt{N_f} \alpha_s/(\pi f_0)~~,
\end{equation}
with $f_0/f_{\pi}\simeq 1$ \cite{ball}, $f_\pi = 93$ MeV being 
the pion decay constant and $\alpha_s=g^2/4 \pi$
the strong coupling constant.
Furthermore, a kinetic term 
$C (\partial_{\mu}K^{\mu})^2$ is added to (\ref{leta}) and the additional 
phenomenological constant $C$ is fitted in order to describe the empirical
$\eta - \eta'$ mass difference.
For a review, see e.g. \cite{miransky}. 

{\bf 2.} In difference to  \cite{crew}-\cite{volkov} we start
from a Lagrangian which includes in addition to (\ref{leta}) the standard
gluon kinetic term
\begin{equation}
\label{leta1}
{\cal L}_{\rm singlet}=
{\cal L}_{\rm singlet}^{\rm meson}- 
\frac{1}{4}G^{\mu\nu a} G_{\mu\nu}^a~.  
\end{equation} 
From the Lagrangian (\ref{leta1})
it should be possible to obtain the mass of the
singlet pseudoscalar $\eta_0$ as a consequence of the coupling to the 
gluon field. To calculate the mass of the  $\eta_0$
let us construct the corresponding Hamiltonian.
We have
\begin{equation} 
G^{\mu\nu a} G_{\mu\nu}^a=-2 (G_{0i}^a)^2 + 2 (B_i^a)^2 ~~,~~
G^{\mu\nu a} \tilde{G}_{\mu\nu}^a=- 4 G_{0i}^a B_i^a~~,
\end{equation}
where $B_i^a$ are the components of the magnetic field strength.
For the quantization of the physical gluon fields $A_0$ has to be eliminated,
we use here the convention $A_0=0$.
Introducing the canonical momenta 

\begin{eqnarray}
\Pi_{\eta_0}&\equiv &
\frac{\partial{\cal L}_{\rm singlet}}
{\partial\dot {\eta}_0}=\dot{\eta}_0~,\nonumber\\  
E_i^a &\equiv &
\frac{\partial{\cal L}_{\rm singlet}}
{\partial\dot {A}_i^a}={G}_{0i}^a+cB_i^a\eta_0~,  
\end{eqnarray} 

the Hamiltonian density reads

\begin{eqnarray}
\label{heta}
{\cal H}_{\rm singlet} &=&
\dot{\eta_0}\Pi_{\eta_0}+\dot {A}_i^a E_i^a - {\cal L}_{\rm singlet}
\nonumber\\
&=&\frac{1}{2}\Pi_{\eta_0}^2 + {1\over 2}(\partial_i\eta_0)^2
 +\frac{1}{2} ({E}_i^a - c B_i^a \eta_0)^2 + \frac{1}{2}(B_i^a)^2~.  
\end{eqnarray} 

We would like to eliminate the gluon degrees of freedom by averaging
the corresponding Hamiltonian $H=\int d^3x{\cal H}$ over the gluon vacuum. 

{\bf 3.} In order to perform the averaging over the nonperturbative gluon 
vacuum we shall use the model of the squeezed gluon vacuum.
The squeezed condensate of gluons has been investigated recently
\cite{celenza}-\cite{fors} in order to construct a Lorentz and gauge
invariant stable QCD vacuum in Minkowski space. 
Different alternative approaches
have not solved this problem. For instance the simple perturbative vacuum 
is unstable \cite{sav}, and there is no stable 
(gauge invariant) coherent vacuum in Minkowski space \cite{leut}.

From the physical point of view, the squeezed state differs from the coherent 
one by the condensation of colour singlet gluon pairs rather than of single 
gluons. In analogy to the Bogoliubov model \cite{nn}
we consider the case of a homogeneous condensate, but
in a squeezed instead of a coherent state.
The squeezed vacuum $|0_{\rm sq}[f_0]>$ 
as a candidate for a homogeneous colourless gluon vacuum is constructed from 
a nonperturbative reference vacuum $|0>~\equiv|0_{\rm sq}[f_0=0]>$, 
further specified below, according to
\begin{equation}
\label{sqt1}
|0_{\rm sq}[f_0]>\ =\ U_{\rm sq}^{-1}[f_0]|0>~.
\end{equation}
The squeezing operator
\begin{equation}
\label{sqt2}
U_{\rm sq}[f_0]= \exp\left[i{f_0\over 2}\sum_{a,i}(A^a_{i}(0)E^a_{i}(0)+
E^a_{i}(0)A^a_{i}(0))\right]
\end{equation}
with the zero momentum components $A^a_{i}(0)$ and $E^a_{i}(0)$ 
of the fields and their canonical momenta contains the  parameter $f_0$ 
given below.
This special transformation for the homogeneous condensate
does not violate Lorentz invariance,
since the gauge fields are massless \cite{Linde}.
The question of gauge invariance is very difficult but as in Ref. 
\cite{schuette} we suppose the gauge invariance of all the spatial zero 
momentum components of the gauge fields.
The multiplicative transformations of fields corresponding to
(\ref{sqt1}) and (\ref{sqt2}) are

\begin{eqnarray}
\label{usq}
U_{\rm sq}[f_0]~A_i^a(0)~U_{\rm sq}^{-1}[f_0] &=& {\rm e}^{ f_0} A_i^a(0)~,
\nonumber\\
U_{\rm sq}[f_0]~E_i^a(0)~U_{\rm sq}^{-1}[f_0] &=& {\rm e}^{ - f_0} E_i^a(0)~.
\end{eqnarray}

After this canonical transformation the squeezed expectation values as
functions of the squeezing parameter $f_0$ behave like

\begin{eqnarray} 
<0_{\rm sq}[f_0]|\left(B_i^a(0)\right)^2|0_{\rm sq}[f_0]> 
&= & e^{4f_0}<0|\left(B_i^a(0)\right)^2|0>~,\label{b2sq}\\
<0_{\rm sq}[f_0]|\left(E_i^a(0)\right)^2|0_{\rm sq}[f_0]> &= &
 e^{-2f_0}<0|\left(E_i^a(0)\right)^2|0>~,\label{e2sq}\\
<0_{\rm sq}[f_0]|E_i^a(0)B_i^a(0)|0_{\rm sq}[f_0]> &= &
 e^{f_0}<0|E_i^a(0)B_i^a(0)|0>~,\label{ebsq}
\end{eqnarray}

which follows from (\ref{usq}), noting that $B^a_i(0)=f^{abc}\epsilon_{ijk}
A^b_j(0)A_k^c(0)$. 
Let the reference vacuum $|0>$ be such that the expectation values
$<0|\left(B_i^a(0)\right)^2|0>$, $<0|\left(E_i^a(0)\right)^2|0>$
and $<0|E_i^a(0)B_i^a(0)|0>$ 
behave in the large volume limit ($V \to \infty$)  like $V^{-4/3}$ in 
accordance with dimensional analysis.
The parameter of the squeezing transformation $f_0$ can be chosen so that
the magnetic condensate density (\ref{b2sq}) remains finite in the large volume
limit ($e^{4f_0}\sim V^{4/3}$). 
This entails that the electric condensate (\ref{e2sq}) and the mixed
condensate (\ref{ebsq}) vanish in this limit 

\begin{eqnarray}
\label{E}
\lim_{V \rightarrow \infty} 
<0_{\rm sq}[f_0]|\left(E_i^a(0)\right)^2|0_{\rm sq}[f_0]>~ &=&
 {\cal O} [1/V^2]  ~,\\
\label{EB}
\lim_{V \rightarrow \infty} 
<0_{\rm sq}[f_0]|E_i^a(0)B_i^a(0)|0_{\rm sq}[f_0]>~ &=&
 {\cal O} [1/V]  ~.
\end{eqnarray}

Hence we conclude that in the squeezed vacuum the gluon condensate
is equal to its magnetic part,

\begin{eqnarray}
\label{asg2}
<\alpha_s G^2>&\equiv& 
<0_{\rm sq}[f_0]|\alpha_s G^{\mu\nu a}G_{\mu\nu}^a|0_{\rm sq}[f_0]> \nonumber\\
&=& 2~ <0_{\rm sq}[f_0]|\alpha_s(B_i^a)^2|0_{\rm sq}[f_0]> ~.
\end{eqnarray}

With these expressions for the averages of the relavant gluon field operators
in hand we shall go on in the next paragraph to
derive an effective Hamiltonian for the $\eta_0$ in the gluon vacuum.

{\bf 4.} Taking the expectation value of the Hamiltonian
corresponding to (\ref{heta}) with respect to  
the squeezed gluon vacuum $|0_{\rm sq}[f_0]>$ we obtain the effective
Hamiltonian 
\begin{equation}
\label{Heta_0}
H_{\rm eff}=\int d^3x\left[
\frac{1}{2}\Pi_{\eta_0}^2 + {1\over 2}(\partial_i\eta_0)^2 + 
\frac{1}{2}\Delta m_0^2\eta_0^2 \right]
 + {\rm const} ~~.
\end{equation}
with the $\eta_0$ mass 
\begin{equation}
\label{meta0}
\Delta m_0^2 =\frac{3 \alpha_s}{\pi^2 f_\pi ^2}
<0_{\rm sq}[f_0]|\alpha_s (B_i^a)^2|0_{\rm sq}[f_0]> ~.
\end{equation}

Note that in the effective $\eta_0$ Hamiltonian (\ref{Heta_0}) 
a term linear in $\eta_0$ does not appear, since it is proportional to 
$<0_{\rm sq}[f_0]|E_i^a B_i^a|0_{\rm sq}[f_0]>$
which according to (\ref{EB}) vanishes in the squeezed vacuum.
Because of (\ref{asg2})
we can rewrite the mass formula (\ref{meta0}) in the form
\begin{equation}
\label{meta0G}
<\alpha_s G^2> =\frac{2 \pi^2 f_\pi ^2}{3 \alpha_s}\Delta m_0^2 ~.
\end{equation}
This formula is the main result of our investigation.
It relates the gluon condensate to the $U_A(1)$ breaking mass shift of the
$\eta_0$. The value of $\alpha_s$ in the low energy region is not known
very well from experiment.
The value used by SVZ \cite{zakharov} is $\alpha_s\approx 1$ and that used
by Narison \cite{narison} in the low energy region is
$\alpha_s(1.3~ {\rm GeV})\simeq 0.64^{+0.36}_{-0.18}\pm 0.02$. 
In order to check whether relation (\ref{meta0G}) is in agreement with
empirical data we have plotted in Fig. 1 the gluon condensate $<\alpha_s G^2>$ 
against $\alpha_s$ for two representative values of $\Delta m_0^2$ 
which are estimates based on the U(3) chiral meson Lagrangian:
The first one is obtained for the  case of vanishing mixing angle and the 
chiral limit, where $\Delta m_0^2=m^2_{\eta'}-m^2_{\eta}= 0.616~ {\rm GeV^2}$.
For the other estimate we use the quadratic Gell-Mann--Okubo mass formula  
and obtain $\Delta m_0^2=m^2_{\eta'}+m^2_{\eta}-2m_K^2=0.729~ {\rm GeV^2}$,
see \cite{volkov}.
\begin{figure}
\centerline{\psfig{figure=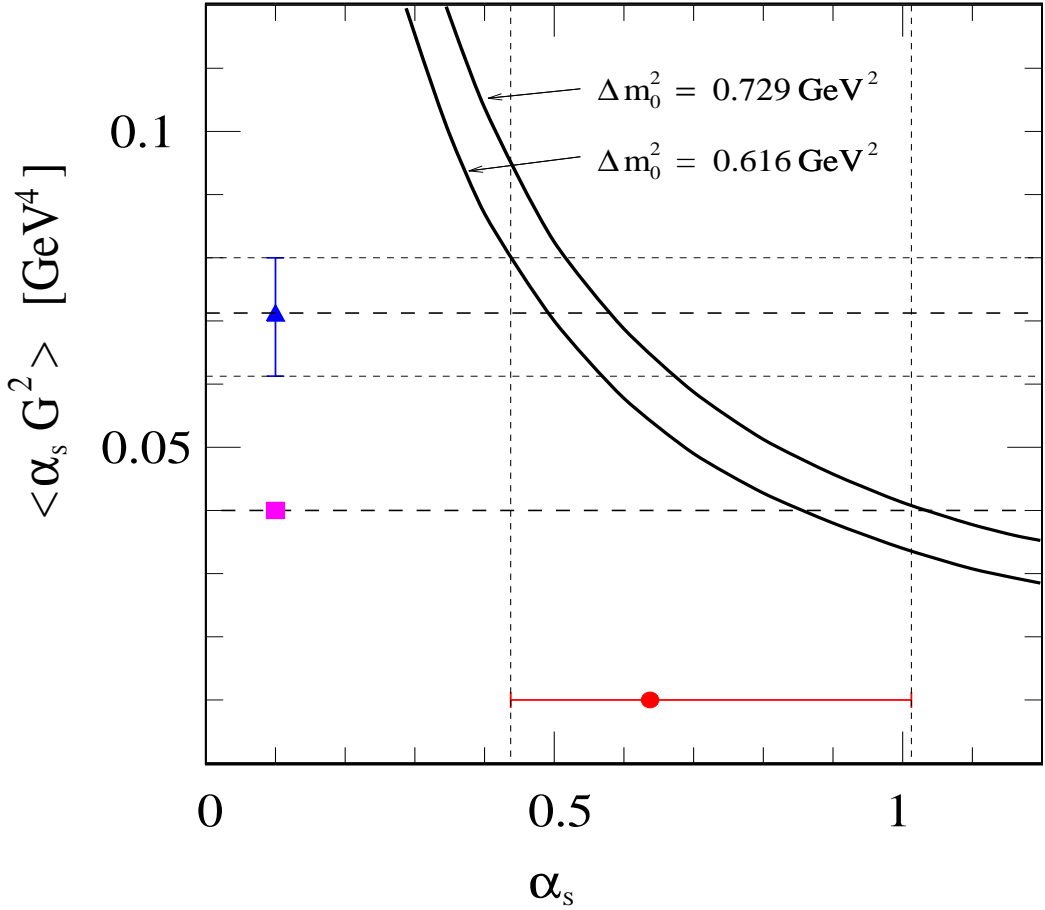,width=14cm,height=12cm,angle=-90}}
\caption{The gluon condensate $<\alpha_s G^2>$ vs. QCD coupling 
$\alpha_s$ for two representative estimates of the $\eta_0$ mass shift
(solid lines) according to (\protect{\ref{meta0G}}).
Also shown are the gluon condensate values obtained by Narison 
\protect{\cite{narison}} (filled triangle) and by Shifman, Vainshtein and 
Zakharov \protect{\cite{zakharov}} (filled square) which both are compatible 
with the $\alpha_s(1.3~ {\rm GeV})$ value (filled circle) of Ref.
\protect{\cite{narison}}. }
\end{figure}
The solid lines in Fig. 1 give our result for these two typical values of
the $\eta_0$ mass shift.
Also shown are the gluon condensate value
$<\alpha_s G^2>\simeq 0.04~ {\rm GeV^4}$ by Shifman, Vainshtein and 
Zakharov \cite{zakharov} (filled square)
and the update average value 
$<\alpha_s G^2>=(0.071\pm 0.009)~ {\rm GeV^4}$ for the gluon condensate 
obtained by Narison \cite{narison} (filled triangle) in a recent
analysis of heavy quarkonia mass-splittings in QCD.
The gluon condensate values described by our result (\ref{meta0G})  
are in good agreement with both the SVZ and the Narison value
for the respective values of $\alpha_s$ in the range of Narison's update
average $\alpha_s$ (filled circle).
This is an interesting and unexpected result of our investigation.
 
{\bf 5.} In conclusion we have considered a mechanism to create a large 
mass for the $\eta_0$ due to its anomalous interaction with the gluons of the 
squeezed vacuum.
In the framework of a squeezed vacuum we have obtained a relation between
the value of the gluon condensate and the mass shift of the $\eta_0$
as a function of the strong coupling constant. 
The gluon condensate values found in our approach for two estimates
of $\eta_0$ mass shifts are in quite good agreement with both the 
``standard'' value $0.04~ {\rm GeV}^4$ by Shifman, Vainsthein and Zakharov 
and the update average value $0.071~ {\rm GeV}^4$ by Narison for reasonable 
values of the strong coupling in the low energy region.
In the present work we have pointed out a special
interesting possibility to resolve the U$_A$(1) problem and to obtain estimates
for the value of the gluon condensate in the framework of
a squeezed gluon vacuum.

\section*{Acknowledgement}
We are grateful to D. Ebert, A.V. Efremov, E.A. Kuraev and 
L.N. Lipatov for fruitful discussions.
HPP is grateful to the Deutsche Forschungsgemeinschaft for support under 
contract No. RO 905/11-1.
The work of VNP and MKV was supported in part by the RFFI, 
grant No. 96-01-01223 and
the Federal Minister for Research and Technology (BMFT) within
Heisenberg-Landau Programme.
Two of us (VNP and MKV) acknowledge the financial support provided
by the Max-Planck-Gesellschaft and the hospitality of the MPG Arbeitsgruppe
''Theoretische Vielteilchenphysik'' at the University of Rostock, where part
of this work has been done.\\
\vspace{5mm}\\
\noindent
Note added:
We wish to thank the referee for his comments and for pointing out 
Ref. \cite{narison} to us.



\begin{thebibliography}{99}  
\bibitem{zakharov}
 M.A. Shifman, A.I. Vainshtein and V.I. Zakharov,
 Nucl. Phys. B {  147} (1979) 385, 448, 519.
\bibitem{narison}
 S. Narison, Phys. Lett. { B 387} (1996) 162.
\bibitem{weinberg}
 S. Weinberg, Phys. Rev. D {  11} (1975) 3583. 
\bibitem{ball}
 P. Ball, J.-M. Fr\`{e}re and M. Tytgat, Phys. Lett. B {  365} (1996) 367.
\bibitem{pham}
 T.N. Pham, Phys. Lett. {  B 246} (1990) 175.
\bibitem{thooft}
 G. 't Hooft, Phys. Rev. D {  14} (1976) 3432; {  18} (1978) 2199 (E) . 
\bibitem{crew}
 R.J. Crewther, Phys. Lett. {  70B} (1977) 349.
\bibitem{veneziano}
 G. Veneziano, Nucl. Phys. {  B 159} (1979) 213, 
 P. Di Vecchia and G. Veneziano, Nucl. Phys. {  B 171} (1980) 253.
\bibitem{rst}
 C. Rosenzweig, J. Schechter and C.G. Trahern, Phys. Rev. D {  21}
 (1980) 3388.
\bibitem{volkov}
 M.K. Volkov, 
 Sov. J. Part. Nuclei {  13} (1982) 446, {  17} (1986) 186.
\bibitem{weise} 
 S. Klimt, M. Lutz, U. Vogl and W. Weise,
            Nucl. Phys. {  A 516} (1990) 429;\\
 S.P. Klevansky, Rev. Mod. Phys. {  64} (1992) 649;\\ 
 T. Hatsuda and T. Kunihiro, Phys. Rep. {  247} (1994) 221.
\bibitem{dmitra}
 V. Dmitrasinovi\'{c}, Phys. Rev. {  C 53} (1996) 1383 and references 
therein. 
\bibitem{miransky}
 V.A. Miransky, Dynamical Symmetry Breaking in Quantum Field Theories
 (World Scientific, Singapore, 1993), ch. 12.
\bibitem{celenza}
 L.S. Celenza and C.M. Shakin, Phys. Rev. D {  34} (1986) 1591.
\bibitem{kogan}
 I.I. Kogan and A. Kovner, Phys. Rev. D {  52} (1995) 3719.
\bibitem{biro}
 T.S. Biro, Ann. Phys. (NY) {  191} (1989) 1; Phys. Lett. B {  278} (1992) 
15; Int. J. Mod. Phys. {  2} (1992) 39.
\bibitem{mishra}
 A. Mishra, H. Mishra, S.P. Misra and S.N. Nayak,
 Phys. Rev. D {  44} (1991) 110;
 Z. Phys. C {  37} (1993) 233;
 A. Mishra, H. Mishra and S.P. Misra, Z. Phys. C {  57} (1993) 241;
 Z. Phys. C {  59} (1993) 159.
\bibitem{fors}
V.N. Pervushin, G. R\"opke, M.K. Volkov, D. Blaschke, H.-P. Pavel, A. Litvin:
Squeezed condensate of Gluons in QCD, Rostock Preprint, MPG-VT-UR 60/96 (1996).
\bibitem{sav}
G.K. Savvidy: Phys. Lett. {  B 71} (1977) 133.
\bibitem{leut}
H. Leutwyler: Nucl. Phys. {  B 179} (1981) 129. 
\bibitem{nn}
 N.N. Bogoliubov, J. Phys. 11 (1947) 23.
\bibitem{Linde} A. Linde: Particle Physics and Inflationary Cosmology
 (Harwood Academic Publishers, Amsterdam, 1990), p. 69.
\bibitem{schuette}
 D. Sch\"utte, Phys. Rev. { D 31} (1985) 810.
\end{thebibliography}
\end{document}